# Spin-Dependent Transport in One-Dimensional Mercury- and Silver-Mediated DNA Nanowires

Arpan De[1*], William Livernois[1], Simon Vecchioni[2], Yoel P. Ohayon[3] and M.P. Anantram[1]

1 Department of Electrical and Computer Engineering, University of Washington, Seattle, WA 98195, USA
2 Department of Life Sciences, Imperial College London, London, SW7 2AZ, UK
3 Department of Chemistry, New York University, New York, NY 10003, USA

*Address correspondence to arpan99@uw.edu

## Abstract:

Metallo-DNA nanowires serve as a versatile, programmable scaffold for molecular-scale electronics. In this work, we demonstrate that periodic heavy-metal coordination fundamentally reshapes the electronic landscape of DNA, generating dispersive, metal-centered conduction bands rather than isolated localized states. Driven by enhanced inter-layer coupling, where second nearest-neighbor hopping plays a crucial role – the conduction bandwidth can approach 1 eV, more than double that of canonical DNA. These broad bands suggest enhanced robustness of transport against structural fluctuations. As a result of relativistic effects, in these quasi-one-dimensional materials, spin-orbit (SO) coupling enhances band dispersion and governs spin-dependent transport. We predict that although ligand hybridization reduces atomic SO splitting, the effective SO splitting in these nanowires is comparable to the next-nearest-neighbor hopping strength, which results in considerable spin mixing. We find that $Hg^{2+}$-coordinated nanowire exhibit pronounced spin-flip conductance comparable to spin-conserving conductance across the LUMO manifold. This efficient spin conversion remains robust over extended molecular lengths and is accompanied by a set of highly delocalized unoccupied states with oscillating nodal structures, readily accessible under low-bias transport measurements. In contrast, the $Ag^{+}$-coordinated nanowires exhibit weak spin-flip scattering that is more than an order of magnitude smaller than spin-preserving transport. This duality establishes the choice of coordination metal as a decisive parameter to govern metal-metal interaction, electron delocalization, and spin-dependent transport. Overall, our findings demonstrate how relativistic effects and quantum confinement intersect in these bio-templated, quasi-one-dimensional molecular conductors.

## Introduction

Over the past three decades, DNA has attracted sustained interest as a quasi-one-dimensional material for electron transport, driven by the fundamental question of whether $\pi - \pi$ stacking between base pairs can enable long-range transport akin to a molecular wire[1,2]. Early theoretical and experimental work focused on Watson-Crick (WC) DNA, debating whether the electronic transport should be understood in terms of metallic[3], polaronic[4,5], semiconducting[6], superconducting[7], coherent[8–10] or hopping[11,12] mechanisms. These efforts sought to probe if DNA has the ability to host electronic states capable of long-range conduction[13–15]. It was found that the energy levels of individual nucleobases differ considerably, inhibiting efficient conduction along the double helix. In particular, adenine-thymine (A:T) segments act as potential barriers relative to guanine-cytosine (G:C) regions, limiting orbital delocalization[16–20]. Ab-initio modeling and tight-binding approaches revealed the nearest neighbor hopping integrals to be at most ~ 100 meV, rendering DNA as a material with marginally delocalized states at best and prone to disorder[21–24]. Further, the application of a bias shows a clear breaking up of the density of states into weakly overlapping regions from base to base, suggesting poor conduction. Consequently, except for rare experimental reports[1,2], electronic transport in native DNA is poor. From a transport standpoint, efficient conduction in quasi-one-dimensional systems requires sufficient orbital delocalization to withstand disorder and structural fluctuations even if the contact Fermi energy is close to band edges. Delocalized states suppress backscattering and enable coherent carrier transport over extended length scales, whereas localized molecular orbitals strongly inhibit ballistic transport. This consideration motivates the search for "post-WC" DNA motifs in which electronic coupling along the helical axis is substantially higher.

Recently, metallo-DNA nanowires have emerged as a promising platform beyond native DNA, in which metal atoms are periodically incorporated into the base-pair stack[25–30]. Metal incorporation is anticipated to enhance orbital overlap and inter-site coupling, providing a compelling motivation to investigate electronic transport in metallo-DNA nanowires. The main distinction of metallo-DNA nanowires from heavy-metal wires[31,32] lies in the unique role of DNA as a scaffold for organizing metal ions into a periodic chain. Through bottom-up self-assembly and metal-mediated base pair chemistry, the metal centers can be incorporated into the DNA structure without requiring a separately synthesized molecular backbone. Moreover, more than one metal ion type can be introduced in a single DNA framework[33]. The established structural stability and programmability of DNA further make it an attractive platform to study low dimensional transport and its potential in spin and charge-based device applications. Despite rapid progress in atomically precise synthesis, the fundamental electronic transport mechanisms in these metalated nanowires remain largely unexplored. In this work, we predict for the first time that the presence of metal-derived orbitals introduces significant spin-flip scattering in DNA. We also show that the metal atoms introduce new conduction channels that have a significant inter-site coupling, facilitating unprecedented orbital delocalization in DNA. We study the electronic transport properties of two distinct, experimentally relevant metallo-DNA nanowires – silver ($Ag^{+}$)- and mercury($Hg^{2+}$)-coordinated – using a two-component relativistic framework, explicitly accounting for spin-orbit coupling arising from the metal centers. Our results elucidate how metal incorporation reshapes the electronic structure and establishes a route toward robust electronic transport beyond the limitations of native DNA.

## Theory

We have considered *the Generalized Kohn-Sham* (GKS) framework, in which each molecular orbital is represented as a two-component spinor formed from the spin basis states $|\alpha\rangle$ and $|\beta\rangle$. The $i^{th}$ GKS spinor is written as: $\Phi_i(r) = \begin{pmatrix} \phi_{i\alpha}(r) \\ \phi_{i\beta}(r) \end{pmatrix} = \phi_{i\alpha}(r)|\alpha\rangle + \phi_{i\beta}(r)|\beta\rangle$, where $\phi_{i\alpha}(r)$ and $\phi_{i\beta}(r)$ are the spin-resolved spatial components. Fully-relativistic pseudopotentials[34] (FRPPs) were used for the metal centers to account for spin-orbit coupling. The GKS Fock operator, $\hat{F}$, is expressed as:

$$\hat{F} = \hat{T}\sigma_0 + \hat{V}_{eff}(r) = \left(-\frac{1}{2}\nabla^2\sigma_0\right) + \left(v^H(r)\sigma_0 + \hat{V}^{XC}(r) + \hat{V}^{ext}(r)\right) \quad (1)$$

where the first term corresponds to the kinetic energy operator ($\hat{T} = -\frac{1}{2}\nabla^2\sigma_0$), $\sigma_0$ is the $2 \times 2$ identity matrix. Here, we have chosen atomic units ($\hbar = m_e = 1$). The Hartree potential, $v^H(r)$, is a scalar potential that acts identically on both spin components. The exchange-correlation potential operator, $\hat{V}^{XC}(r)$, and the spin-dependent external potential operator, $\hat{V}^{ext}(r)$, are $2 \times 2$ matrices in spin space. The external potential operator includes the electron-nucleus interaction ($v^{Ne}(r)$) and the relativistic pseudopotential operator ($\hat{V}^{PP}(r)$). Within the fully relativistic pseudopotential framework, $\hat{V}^{PP}(r)$ consists of spin-free average scalar-relativistic (SR) and spin-dependent (SO) terms. For a given orbital with angular momentum $l$, these two terms denoted by $V_l^{SR}(r)$ and $V_l^{SO}(r)$ respectively, are defined as, $V_l^{SR}(r) = \frac{1}{2l+1}\left[(l+1)V_{l,j=l+\frac{1}{2}}(r) + (l)V_{l,j=l-\frac{1}{2}}(r)\right]$ and $V_l^{SO}(r) = \frac{2}{2l+1}\left[V_{l,j=l+\frac{1}{2}}(r) - V_{l,j=l-\frac{1}{2}}(r)\right]$. Here, $V_{l,j=l+\frac{1}{2}}(r)$ and $V_{l,j=l-\frac{1}{2}}(r)$ are the relativistic pseudopotentials associated with the states with total angular-momentum quantum numbers $j = l + \frac{1}{2}$ and $j = l - \frac{1}{2}$ respectively. Therefore, the fully relativistic pseudopotential operator, denoted by $\hat{V}^{PP}(r)$, can be expressed as[35–38]:

$$\hat{V}^{PP}(r) = \sum_{l,m} |l,m\rangle\left[V_l^{SR}(r)\sigma_0 + V_l^{SO}(r)\,\hat{L}\cdot\hat{S}\right]\langle l,m| \quad (2)$$

where $\hat{L} = (\hat{L}_x, \hat{L}_y, \hat{L}_z)$ and $\hat{S} = (\hat{S}_x, \hat{S}_y, \hat{S}_z)$ are orbital angular momentum operator and spin operator respectively; $|l,m\rangle\langle l,m|$ is the projector operator onto the state with quantum numbers $l$ and $m$. Equation (2) in spin-blocked form can be expressed as:

$$\begin{aligned}\hat{V}^{PP}(r) &= \frac{1}{2}\begin{pmatrix} \sum_{l,m}|l,m\rangle\left[2V_l^{SR}(r) + V_l^{SO}(r)\hat{L}_z\right]\langle l,m| & \sum_{l,m}|l,m\rangle\left[V_l^{SO}(r)(\hat{L}_x - i\hat{L}_y)\right]\langle l,m| \\ \sum_{l,m}|l,m\rangle\left[V_l^{SO}(r)(\hat{L}_x + i\hat{L}_y)\right]\langle l,m| & \sum_{l,m}|l,m\rangle\left[2V_l^{SR}(r) - V_l^{SO}(r)\hat{L}_z\right]\langle l,m| \end{pmatrix} \\ &= \begin{pmatrix} v_{\alpha\alpha}^{PP}(r) & v_{\alpha\beta}^{PP}(r) \\ v_{\beta\alpha}^{PP}(r) & v_{\beta\beta}^{PP}(r) \end{pmatrix}\end{aligned} \quad (3)$$

The GKS Fock-operator in its full glory can be written as:

$$\begin{aligned}\hat{F} &= \begin{pmatrix} -\frac{1}{2}\nabla^2 + v^H(r) + v_{\alpha\alpha}^{ext}(r) + v_{\alpha\alpha}^{XC}(r) & v_{\alpha\beta}^{ext}(r) + v_{\alpha\beta}^{XC}(r) \\ v_{\beta\alpha}^{ext}(r) + v_{\beta\alpha}^{XC}(r) & -\frac{1}{2}\nabla^2 + v^H(r) + v_{\beta\beta}^{ext}(r) + v_{\beta\beta}^{XC}(r) \end{pmatrix} \\ &= \begin{pmatrix} \hat{F}_{\alpha\alpha} & \hat{F}_{\alpha\beta} \\ \hat{F}_{\beta\alpha} & \hat{F}_{\beta\beta} \end{pmatrix}\end{aligned} \quad (4)$$

where the spin-resolved Hartree potential is $v_{\sigma\sigma'}^H(r) = \delta_{\sigma\sigma'}\int\frac{n(r')}{|r-r'|}d^3r' = \delta_{\sigma\sigma'}v^H(r)$ with $\sigma, \sigma' = \alpha, \beta$ and $n(r)$ is the total electron density[39]. Because the Hartree potential is spin independent, $v_{\alpha\beta}^H(r) = v_{\beta\alpha}^H(r) = 0$, and $v_{\alpha\alpha}^H(r) = v_{\beta\beta}^H(r) = v^H(r)$. The spin-dependent external potential is given by $v_{\sigma\sigma'}^{ext}(r) = v^{Ne}(r)\,\delta_{\sigma\sigma'} + v_{\sigma\sigma'}^{PP}(r)$. $v_{\sigma\sigma'}^{XC}(r)$ is the spin-dependent exchange-correlation potential. The one-particle GKS equation is given by: $\hat{F}\Phi_i = \epsilon_i\Phi_i$. In the atomic-orbital basis, each spatial orbital can be expanded in terms of the basis functions as: $\phi_{i\alpha(i\beta)}(r) = \sum_\nu C_{\nu\alpha(\nu\beta)}^i\chi_\nu(r)$, where $\chi_\nu(r)$ is the $\nu^{th}$ basis function and $C_{\nu\alpha}^i, C_{\nu\beta}^i$ are the expansion coefficients for the two spin components of the $i^{th}$ spinor. Thus, the GKS equation in the non-orthogonal basis can be written as:

$$\begin{pmatrix} \mathbb{F}_{\alpha\alpha} & \mathbb{F}_{\alpha\beta} \\ \mathbb{F}_{\beta\alpha} & \mathbb{F}_{\beta\beta} \end{pmatrix}\begin{pmatrix} \mathbb{C}_{i\alpha} \\ \mathbb{C}_{i\beta} \end{pmatrix} = \epsilon_i\begin{pmatrix} \mathbb{S}_{\alpha\alpha} & 0 \\ 0 & \mathbb{S}_{\beta\beta} \end{pmatrix}\begin{pmatrix} \mathbb{C}_{i\alpha} \\ \mathbb{C}_{i\beta} \end{pmatrix} \rightarrow \mathbb{F}\mathbb{C} = \epsilon_i\mathbb{S}\mathbb{C} \quad (5)$$

where $\mathbb{F}_{\sigma\sigma'}^{\mu\nu} = \int\chi_\mu^*(r)\hat{F}_{\sigma\sigma'}\chi_\nu(r)d^3r$, $\mathbb{S}_{\sigma\sigma}^{\mu\nu} = \int\chi_\mu^*(r)\chi_\nu(r)d^3r$, and $\mathbb{C}_{i\sigma} = \left(C_{1\sigma}^i\ C_{2\sigma}^i \ldots\right)^T$.

## Computational Methodology

### A. Structure generation:

We have modeled *four* distinct DNA motifs. First, a 13 base-pair long mismatched poly-(CC) and poly-(TT) duplexes with a rise of 3.4 Å and a twist of 36° between two consecutive base pair layers was generated in conventional B-form[40]. These mismatched structures are labelled as non-metalated *C13* and non-metalated *T13* respectively. From these structures, the coordinates of the two N3 atoms within each (C-C) or (T-T) base pair were extracted, and a single $Ag^{+}$ or $Hg^{2+}$ ion was placed at their geometric center to form the corresponding metal-mediated base pairs, as it has been demonstrated that linearly-juxtaposed N3 atoms are the primary coordination site for these metal pairs[25]. This yielded the 13 base-pair long metallo-DNA nanowires, labeled as *C13* and *T13* $(C:Ag^{+}:C)_{13}$ and $(T:Hg^{2+}:T)_{13}$) respectively. Atomic representations of all four structures are provided in Figure S1.

From Protein Databank (PDB) entries 5IX7[28], 7XKM[41], we extracted the N3-*Ag* distance in a $C:Ag^{+}:C$ metalated base pair to range between 2.03-2.36 Å. Similarly, from PDB entries 7SD6[25], 7TG6, 6IUE[29], we extracted the N3-*Hg* distance in a $T:Hg^{2+}:T$ metalated base pair to range between 2.10-2.40 Å. Moreover, J. Kondo *et al*. have reported the N3-*Hg* distance to vary between 2.20-2.40 Å[42]. In our metallo-DNA structures, the distance between N3-$Ag^{+}$ in each layer of *C13* is 2.29 Å, while the N3-*Hg* distance in each layer of *T13* is 2.35 Å. The distance between consecutive metals in both *C13* and *T13* is 3.4 Å. The metal-N3 distances in our recent molecular dynamics study, in which we developed a custom force field for metalated base pairs was found to be comparable at 2.20-2.23 Å and 2.34-2.45 Å for $Ag^{+}$ and $Hg^{2+}$ ions respectively[43]. All remaining base-pair and helical parameters were taken directly from the canonical B-DNA geometry generated by NAB[40].

## B. Ab-initio DFT Modeling:

Density functional theory (DFT) calculations were carried out with Gaussian[44] using the CAM-B3LYP hybrid exchange-correlation functional with the 6-31G** basis set for non-metal atoms and the dhf-SVP-2c basis set for metal atoms, along with fully-relativistic pseudopotentials[34]. 6-31G** basis set for DNA atoms was previously found to yield similar trends as larger basis sets[45] in dsDNA conductance calculations[46,47]. The metal basis sets were chosen based on the spin-orbit splitting benchmarks in Table S1. We have also shown a comparison between non-relativistic and scalar-relativistic energy levels in Figure S2. CAM-B3LYP functional has also been found to effectively model heavy metals, including relativistic effects[48–50]. The polarizable continuum model (PCM) was included to account for the solvent effect. All self-consistent field (SCF) calculations use default 'tight' convergence criteria, requiring the root-mean-square (RMS) and maximum changes in the density matrix between successive iterations to be less than $(10^{-8})$ and $(10^{-6})$, respectively. Each 13 base-pair DNA duplex contains 24 phosphate groups, resulting in a net backbone charge of (-24$e$) as counterions are not included. The total charges for *C13* and *T13* nanowires are set to be (-11$e$) and (+2$e$), respectively, while the non-metalated counterparts have a charge of (-24$e$).

To work with an orthogonal atomic basis set, the system Hamiltonian ($H_{DNA}$) was constructed from the Fock ($\mathbb{F}$) and Overlap ($\mathbb{S}$) matrices by applying a Lowdin transformation[51,52], as shown in the equation below:

$$H_{DNA} = \mathbb{S}^{-\frac{1}{2}}\mathbb{F}\mathbb{S}^{-\frac{1}{2}} \tag{6}$$

The diagonal elements of the Hamiltonian represent the onsite energies of the orbitals, while the off-diagonal elements correspond to the hopping energies between orbitals.

## C. Transport and Density of States Calculations:

For charge transport calculations, we first compute the retarded Green's function ($G^r$), incorporating the self-energies of the contacts as given by the following equation[20]:

$$[E\,\mathbb{S} - \mathbb{F} - \Sigma_L - \Sigma_R]G^r = I \tag{7}$$

where $E$ is the energy, $\Sigma_{L(R)}$ $(= -\frac{i\Gamma_{L(R)}}{2})$ are the self-energies due to left (right) contacts, while $\Gamma_{L(R)}$ represent the corresponding coupling between DNA and left (right) contacts. We have considered wide-band limit for the contacts, meaning the contact self-energies are purely imaginary and energy-independent, Further, $\Gamma_{L(R)}$ are diagonal matrices coupled to all orbitals belonging to the terminal base pairs at the two ends of the nanowire. The coherent transmission ($T$) as a function of energy is expressed as:

$$T = trace(\Gamma_L G^r \Gamma_R G^a) \tag{8}$$

where $G^a = (G^r)^\dagger$ is the advanced Green's function. We have used $\Gamma_{L/R} = 1$ eV. The conductance as a function of the Fermi energy $E_F$ is calculated using $G(E_F) = \int T(E)\left(-\frac{\partial f}{\partial E}\right) dE$, where $f = \frac{1}{1+\exp\left(\frac{E-E_F}{k_B T}\right)}$ is the Fermi function. The conductance is calculated at a temperature ($T$) of 298 K.

The partial density of states (PDOS) at an energy $E$ is calculated for the bare nanowire without contact broadening. Similar to equation (7), we solve the retarded Green's function ($G^r_{mol}$) without contact self-energies as follows:

$$[(E + i\eta)\mathbb{S} - \mathbb{F}]G^r_{mol} = I \tag{9}$$

where, $\eta = 0.01$ is a small imaginary broadening. Next, the partial density of states (PDOS) for $i^{th}$ orbital at energy $E$ is computed as:

$$PDOS(i, E) = -\frac{1}{\pi} Im\left(G^{r'}_{mol}(i, i, E)\right) \tag{10}$$

where, $G^{r'}_{mol} = (G^r_{mol} * \mathbb{S})$. The probability of finding the electron in layer $k$ corresponding to eigenfunction $\Psi$ is $\sum_{j=1}^{N_k} |\Psi_j|^2$, where $j$ is the orbital index in layer $k$ and $N_k$ is the total number of spin orbitals in the $k^{th}$ layer, and the probability density to find the electron on the $i$-th metal ($M_i$) at eigenfunction $\Psi$ is $\sum_{j=1}^{N_{M_i}} |\Psi_j|^2$, where $N_{M_i}$ is the total number of spin orbitals on a metal atom $M_i$ and the orbitals on the $i$-th metal is represented by index $j$.

Our treatment does not explicitly include decoherence or electron-phonon coupling and counterions. Incorporating environmental dephasing could modify spin dynamics and lead to broadening of conduction channels[20,52]. Thus, the conductance reported here represents the phase-coherent benchmark for these metallo-DNA nanowires. Moreover, our previous study showed that, in an aqueous environment, the electrostatic effects of sodium counterions are attenuated by solvent screening and therefore they do not significantly impact transport through the frontier orbitals[45]. Here, we have verified through molecular dynamics simulations[43] that chlorine counterions are at a broad distribution around 0.5 nm from the phosphate backbone. Therefore, these counterions are expected to act as long-range electrostatic potentials without forming bonds with the metalated DNA duplex. Hence, we posit the effect of counterions in altering transport properties of metallo-DNA nanowire will be small. Finally, the present study does not explicitly consider the effect of thermal disorder or conformation fluctuations. We note that previous experimental and theoretical studies have demonstrated that metal-mediated base pairs possess higher binding energies and thermodynamic stability than canonical hydrogen-bonded base pairs (A:T and G:C)[53–55]. This enhanced structural rigidity indicates that metallo-DNA nanowires experience lower local fluctuations in solution at the metalated layer compared to native DNA as demonstrated by Livernois *et al.*[43]

### D. Inter-base-pair Hopping Energy Extraction:

To extract the hopping energies between base-pairs in the $N$ base pair long metallo-DNA nanowire, we first perform a unitary transformation on the system Hamiltonian (equation 6):

$$H^0_{DNA} = \mathbb{U}^\dagger H_{DNA} \mathbb{U} \tag{11}$$

where the block-diagonal unitary matrix $\mathbb{U}$ is defined as:

$$\mathbb{U} = \begin{pmatrix} \mathbb{U}_1 & 0 & \cdots & 0 \\ 0 & \mathbb{U}_2 & \cdots & 0 \\ \vdots & \vdots & \ddots & \vdots \\ 0 & 0 & \cdots & \mathbb{U}_N \end{pmatrix} \tag{12}$$

Here, $U_k$ ($k \in [1, N]$) contains the eigenvectors of the sub-matrix corresponding to the $k^{th}$ base-pair layer of $H_{DNA}$. The base-pair contains the coordinated metal and the two nucleobases that it is bonded to. Consequently, the transformed hopping matrix $t_{ij}$ between the $i^{th}$ and $j^{th}$ base-pairs is given by:

$$t_{ij} = \mathbb{U}_i^\dagger H_{ij} \mathbb{U}_j \tag{13}$$

where $H_{ij}$ denotes the off-diagonal sub-matrix coupling base pairs $i$ and $j$ in $H_{DNA}$. Next, we calculate the average nearest-neighbor ($t_{p,q}^{1NN}$) and next-nearest-neighbor ($t_{p,q}^{2NN}$) hopping energies across all base-pairs for transition from eigenstate $p$ on one base pair to $q$ on another:

$$t_{p,q}^{1NN} = \frac{1}{N-1}\sum_{w=1}^{(N-1)} \left|t_{w,w+1}(p,q)\right|$$
$$t_{p,q}^{2NN} = \frac{1}{N-2}\sum_{w=1}^{(N-2)} \left|t_{w,w+2}(p,q)\right| \tag{14}$$

In the study, we compute the inter-base-pair hopping energies $\left(t_{p,q}^{1NN}, t_{p,q}^{2NN}\right)$ across the key frontier orbitals transitions: $(p,q) \in \{\mathrm{HOMO}, \mathrm{LUMO}, \mathrm{LUMO}+1, \mathrm{LUMO}+2\}$ of the individual base pairs. Partitioning the full system Hamiltonian $H_{DNA}$ into localized sub-blocks can induce unphysical spurious energy levels within the HOMO-LUMO gap of the whole nanowire. To prevent choosing these artificial states, the local HOMO (LUMO) of individual base pair is identified as the eigenstate whose eigenvalue lies closest to the global HOMO (LUMO) energy level of the whole nanowire.

## Results and Discussion

In this section, we discuss (i) the electronic properties of metallo-DNA nanowires, (ii) the influence of coordinated metal species ($Ag^{+}$ *vs* $Hg^{2+}$) and (iii) spin-flip scattering mechanism in these nanowires.

### A. Charge transport properties:

The schematic representation of a thirteen-base-pair (bp) metallo-DNA nanowire is illustrated in Figure 1(a). We have investigated a prototype metallo-DNA nanowire, *T13* or *(T:$Hg^{2+}$:T)$_{13}$*, comprising 13 repeating units of *T:$Hg^{2+}$:T* metalated base pairs. To gain an understanding of conduction in these metallo-DNA nanowires, we begin by analyzing the spectrum of allowed energy levels, as shown in Figure 1(a, *bottom*). Our calculations reveal the energy gap in *T13* to be 3.13 eV lower than that of their non-metalated (NM) counterparts. The decrease in the energy gap is due to metal-mediated unoccupied energy levels inside the intrinsic gap of the DNA that are absent in the corresponding non-metalated DNA. To further elucidate the role of metal ions in controlling the lowest unoccupied energy levels, we have computed the probability density of low-lying unoccupied eigenfunctions at metal atoms ($|\Psi_{metal}|^2$) as depicted in Figure S3. Indeed, we observe the metals to contribute significantly to the lowest unoccupied energy band. Specifically, the thirteen $Hg^{2+}$ ions in *T13* contribute 42% at the LUMO state. Moreover, $|\Psi_{metal}|^2$ in *T13* does not become smaller than 0.37 at any energy within the frontal unoccupied states. This establishes the fact that the additional unoccupied energies within the intrinsic energy gap of DNA are mainly induced by the $Hg^{2+}$ coordination. Additionally, we confirm that the metal contribution near the highest occupied energy level is very low ($\leq$ 0.1%) for the metalated nanowire. The role of metal in electron flow is further substantiated by the partial density of states (PDOS), as shown in Figure 1(b). Firstly, the overall DOS is considerably higher in the metalated nanowires, compared to non-metalated ones, over an extended energy window (Figure 1(b, *black* and *red lines*)). Additionally, the partial DOS (PDOS) on the metal chain in Figure 1(b, *gray line*) shows that the total DOS on just 13 metal ions is comparable to the total DOS on $\approx$ 800 atoms in the DNA in the LUMO band. It was also found that PDOS on metals in *T13* accounts for 36-44% of the total DOS in the frontal unoccupied bands, signifying the presence of an extended metal-mediated pathway for quantum transport. We have also studied the PDOS at each metalated base pair. Our analysis compares the DOS with and without metal atoms in the energy window from LUMO-1 to LUMO+1 eV (see Figure 1(c)). The non-metalated nanowire in Figure 1(c, *left*) can be characterized with two features – localized 1$^{st}$ and 2$^{nd}$ LUMO energies at terminal base pairs and a high DOS energy window positioned slightly away from the LUMO with an approximate width of 260 meV. In contrast, upon metalation, the energy spectrum changes drastically. As the entire lowest unoccupied band is predominantly contributed by the $Hg^{2+}$ chain, the localized LUMO energies disappear and instead there are multiple highly delocalized *near-LUMO quantized states* in *T13*, starting with the LUMO itself (Figure 1(c, *right*). Further, the spatially resolved partial density of states has $(n-1)$ nodes for the $n^{th}$ unoccupied energy level, akin to a particle-in-a-box induced by the $Hg^{2+}$ coordinated metallo-DNA nanowire (Figure 1(d)).

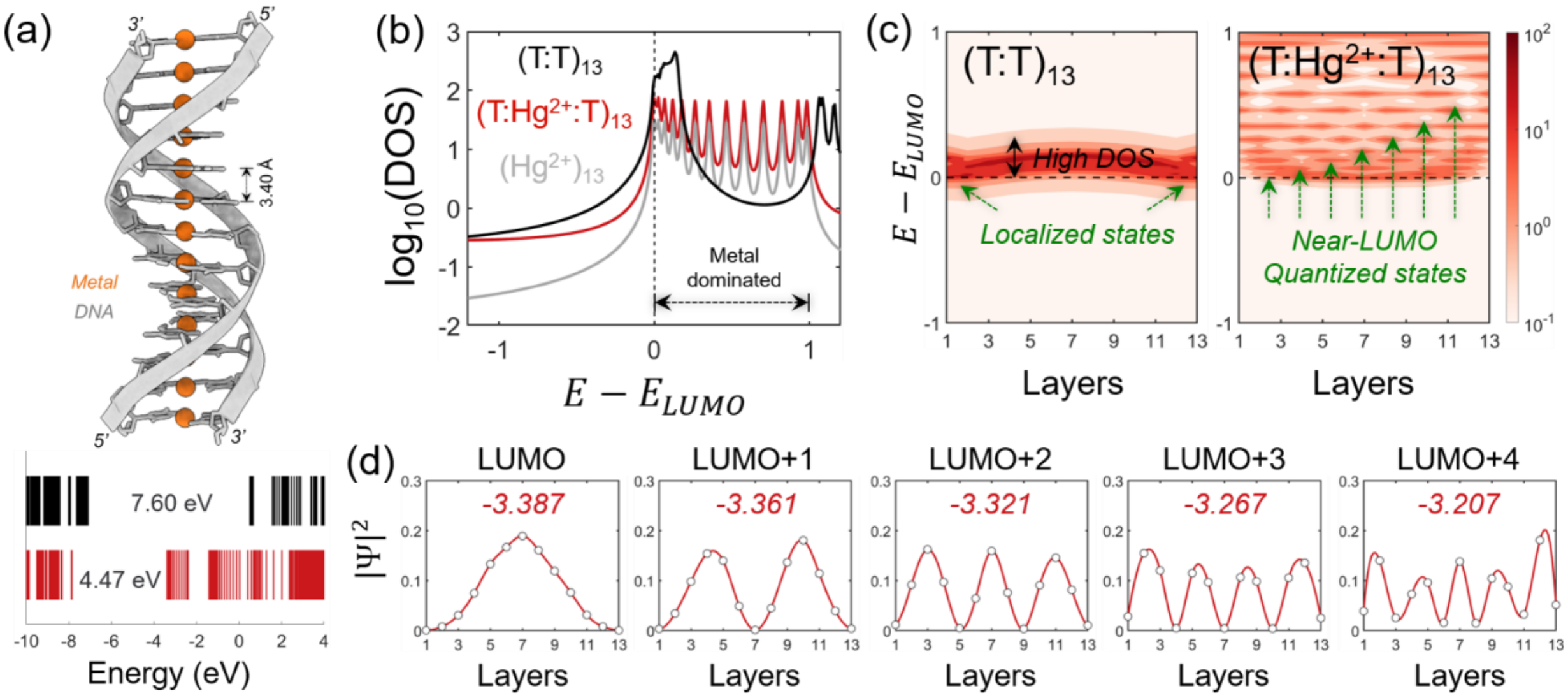

**Figure 1:** (a) (*top*) Schematic of the 13-bp metallo-DNA nanowire, comprising 13 homogenous metalated base pairs. DNA atoms are shown in *gray*, while coordinated metals are depicted in *orange*. (*bottom*) The allowed energy eigenvalues of $(T:Hg^{2+}:T)_{13}$ nanowire and its non-metalated counterpart $(T:T)_{13}$ are shown in *red* and *black* lines respectively. Panel (b) shows the density of states (DOS) as a function of energy for $(T:Hg^{2+}:T)_{13}$. The *black line* corresponds to non-metalated nanowire, $(T:T)_{13}$, while the *red line* represents metalated-nanowire, $(T:Hg^{2+}:T)_{13}$. The partial density of states (PDOS) of the thirteen metal atoms $((Hg^{2+})_{13})$ in the metallo-DNA nanowire is highlighted in *gray*. Panel (c) shows the base-pair (layer)-resolved partial density of states (PDOS) for $(T:Hg^{2+}:T)_{13}$. For non-metalated nanowires (c, *left*), the localized LUMO states are indicated with *green* arrows. In metalated cases (c, *right*), near-LUMO quantized states are highlighted with *green* arrows. (d) Layer-wise contribution in the first five LUMOs of $(T:Hg^{2+}:T)_{13}$ nanowires. The numbers in each subplot represent the corresponding energies in eV. The *red line* is the spline-fitting to the raw data shown in *white empty circles*. In (c, d), a single layer comprises two nucleotide bases and one coordinated metal ion. In the main text, $(T:Hg^{2+}:T)_{13}$ has also been referred to as *T13*.

Additionally, the energy spacing between two successive unoccupied energy levels in *T13* varies between 26-122 meV, making these states individually accessible in experiments.

## B. Mercury ($Hg^{2+}$) *vs* Silver ($Ag^{+}$) coordinated nanowires:

The coordination chemistry of silver with DNA has been extensively studied[25,27,41,56–58]. Preliminary advances include the demonstration of silver-DNA nanowires, *C*-$Ag^{+}$-*C* nanorings formed from kilobase-long polycytosine strands, highly ordered $Ag^{+}$-coordinated duplex nanowires extending from micrometers to millimeters, and even $Ag^{+}$-DNA rods containing no Watson-Crick base pairs. At the same time, silver and mercury ions differ in atomic number, DNA binding strength, orbital character, and spin-orbit coupling strength. Comparing nanowires coordinated with these two metals therefore enables us to assess the tunability of spin-dependent transport properties in these metaled motifs. Akin to *T13*, we construct a nanowire consisting of 13 repeating $C:Ag^{+}:C$ metalated base pairs. Throughout this work, this motif is denoted as *C13* or $(C:Ag^{+}:C)_{13}$.

Consistent with *T13*, the silver atoms exhibit a dominant contribution to the LUMO manifold in *C13*. However, in *C13*, metal-induced energy levels do not exist deep inside the bandgap of the non-metalated DNA, resulting in a modest bandgap reduction of ~ 0.5 eV relative to the non-metalated motif (Figure S4). In Figure 2(a, c), we present the linear-response conductance of the two nanowires around their respective LUMO energies. In both systems, pronounced conductance peaks emerge at the metal-induced unoccupied energy levels, demonstrating the ability of these states to support efficient charge transport. Another notable feature of the metalated nanowires is the emergence of significantly broadened LUMO bands. The bandwidth in *T13* approaches 1 eV in both systems, representing a substantial enhancement over the ~ 0.43 eV bandwidth reported for canonical DNA.[23] In both *C13* and *T13*, the width of the lowest unoccupied states is correlated with the metal contribution, $|\Psi_{metal}|^2$. However, an important feature in *C13* is the decrease in $|\Psi_{metal}|^2$ rapidly within 0.12 eV above the LUMO, producing a dip in conductance profile and a relatively narrow first LUMO band. However, at higher energies, the metal contributions remain significant, resulting in a much wider second unoccupied band (Figure 2(c)). Similarly, in *T13*, the larger metal contribution to the first few unoccupied states produces a broad first LUMO band (Figure 2(a)). To gain further insight into the broadened conduction bands, we next examine the effective tight-binding (hopping) parameters for each of the Kramers doublet extracted from the DFT energy spectrum (Table 1). Strong electronic coupling is observed both within and between the lowest three unoccupied states. The intra-LUMO couplings $(|t^L_{kk}|, k = \{1,2,3\})$ are accompanied by considerable inter-LUMO couplings, $\left(|t^L_{kk'}|, k' > k,\ k, k' \in \{1,2,3\}\right)$. Notably, the $Hg^{2+}$-mediated LUMO state in *T13* exhibits a nearest-neighbor hopping strength of 146.1 meV, approximately 5.6 times more than the corresponding value of 26.1 meV for the $Ag^{+}$-mediated state in *C13*. The importance of metal coordination is further seen in the markedly enhanced second-nearest-neighbor couplings relative to the non-metalated nanowires. Although *T13* has the larger nearest-neighbor LUMO coupling, *C13* exhibits broader unoccupied bands. This difference arises from the energy spacing between the frontier orbitals. In *T13*, the energy separation between the LUMO and LUMO+1 of each metalated base pair is approximately 1.66 eV, indicating that the lowest conduction band is formed mainly from LUMO-derived states of individual metalated base pair (Figure 2(a)). However, in *C13*, the LUMO and LUMO+1 of each metalated base pair are separated by only about 40 meV, allowing strong hybridization between the two orbitals. As a result, both orbitals in each metalated base pair contribute to the first conduction band (Figure 2(c)). The LUMO+2 state, located about 0.9 eV above the LUMO+1, also exhibits a large coupling strength of 180.8 meV. This strong coupling produces the highly dispersed second conduction band in . Further hybridization with lower-lying unoccupied states broadens the overall LUMO-derived manifold. Therefore, the conduction-band widths are governed not only by the magnitudes of the individual hopping integrals, but also by the energy separation and hybridization among the unoccupied orbitals. The resulting broad and strongly interconnected LUMO manifolds are expected to exhibit enhanced robustness against structural and environmental perturbations, making metal-mediated DNA nanowires particularly attractive for bioelectronic applications. In contrast to the enhanced electron transport observed in the LUMO regime, conduction-band widths are governed not only by the magnitudes of the individual hopping integrals, but also by the energy separation and hybridization among the unoccupied orbitals. The resulting broad and strongly interconnected LUMO manifolds are expected to exhibit enhanced robustness against structural and environmental perturbations, making metal-mediated DNA nanowires particularly attractive for bioelectronic applications. In contrast to the enhanced electron transport observed in the LUMO regime, an important result is that both *C13* and *T13* display suppressed conductance in the HOMO regime (Figure 2(b, d)). Metal coordination disrupts the hole transport pathways present in the non-metalated nanowires, leading to much weaker HOMO coupling. The nearest-neighbor HOMO hopping integrals are reduced to only 0.2 and 0.8 meV in *C13* and *T13*, nearly two orders of magnitude smaller than those of their non-metalated counterparts (Table 1). Thus, metal coordination simultaneously suppresses HOMO-mediated hole transport and enhances LUMO-mediated electron transport, shifts the dominant transport mechanism from hole conduction (HOMO-mediated) to electron conduction (LUMO-mediated).

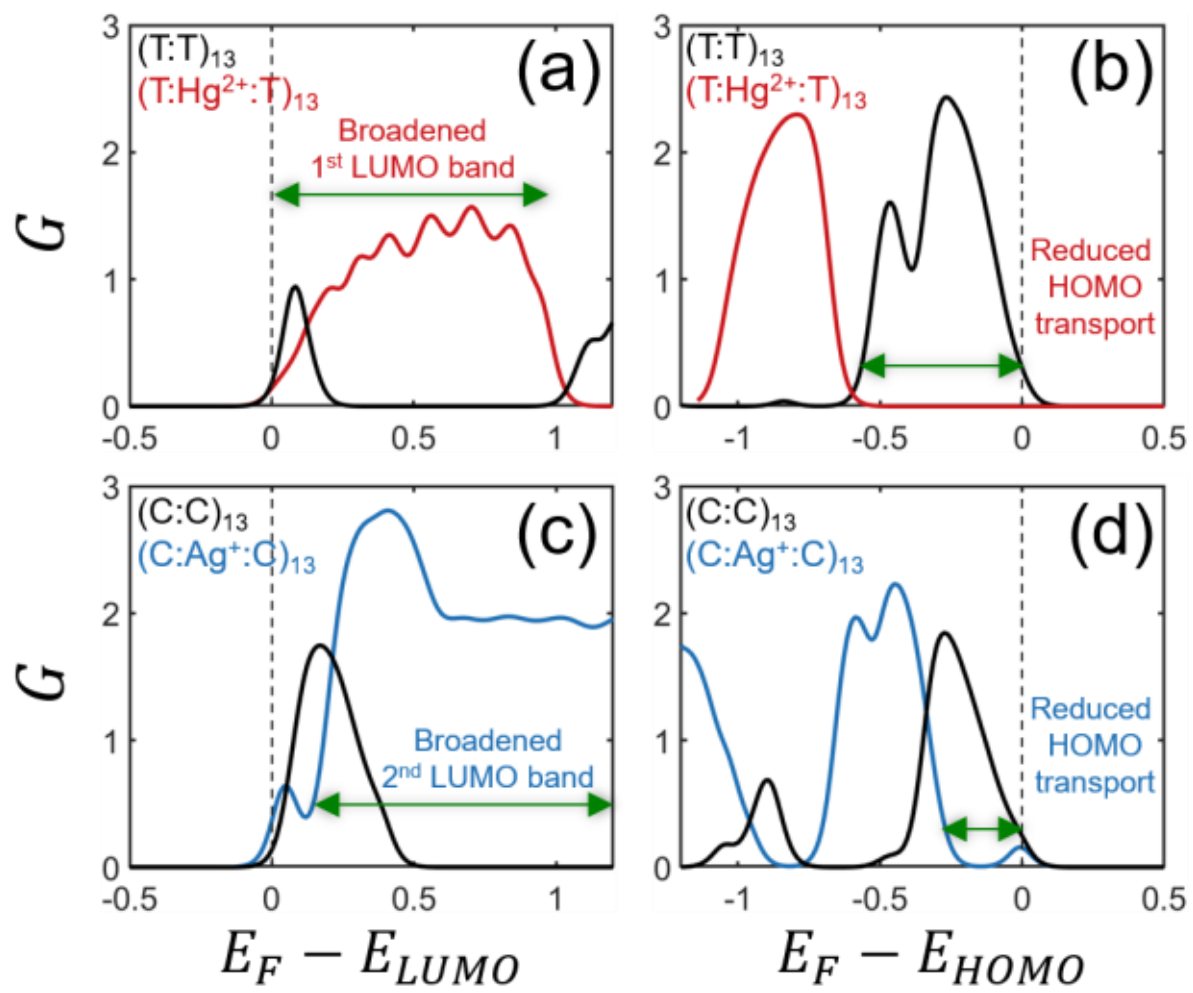

**Figure 2:** Linear-response conductance of metallo-DNA nanowires - (a, b) (T:Hg$^{2+}$:T)$_{13}$ and (c, d) (C:Ag$^{+}$:C)$_{13}$. The *left* and *right* columns correspond to the LUMO and HOMO transport channels, respectively. In each panel, the *colored* curves represent the metal-coordinated systems obtained from two-component relativistic calculations, while the *black* curves denote the corresponding canonical DNA counterparts (T:T)$_{13}$ and (C:C)$_{13}$. Metal coordination results in enhanced LUMO transport while HOMO conduction is suppressed. Here, the conductance $(G)$ is in units of $q^2/h$. The corresponding figures on the *log scale* are provided in Figure S5. In the main text, (T:Hg$^{2+}$:T)$_{13}$ and (C:Ag$^{+}$:C)$_{13}$ have also been referred to as *T13* and *C13*.

**Table 1.** Average inter-base-pair hopping parameters (in meV) for each of the Kramers doublets at HOMO ($|t^H|$) and LUMO ($|t^L|$) states. $|t^L_{\mu\nu}|$ corresponds to hopping between (LUMO+$\nu$) of one base pair to (LUMO+$\mu$) of another base pair. $\Delta^L_{\mu\nu}$ denotes average energy difference between (LUMO+$\nu$) and (LUMO+$\mu$) eigenvalues (in eV) across all base pairs. All the hopping values are rounded off to one decimal place.

| **Energy difference (eV)** | | **1st Nearest Neighbor (⟨..⟩) (meV)** | | | | | | | **2nd Nearest Neighbor (⟨⟨..⟩⟩) (meV)** | | | | | | |
|---|---|---|---|---|---|---|---|---|---|---|---|---|---|---|---|
| $\Delta^L_{10}$ | $\Delta^L_{21}$ | $\|t^H_{00}\|$ | $\|t^L_{00}\|$ | $\|t^L_{11}\|$ | $\|t^L_{22}\|$ | $\|t^L_{01}\|$ | $\|t^L_{12}\|$ | $\|t^L_{02}\|$ | $\|t^H_{00}\|$ | $\|t^L_{00}\|$ | $\|t^L_{11}\|$ | $\|t^L_{22}\|$ | $\|t^L_{01}\|$ | $\|t^L_{12}\|$ | $\|t^L_{02}\|$ |
| **(T:T)$_{13}$** (or Non-metalated *T13*) | | | | | | | | | | | | | | | |
| 0.01 | 1.39 | 77.7 | 15.9 | 16.5 | 32.7 | 3.3 | 38.3 | 55.4 | 1.0 | 0.2 | 0.3 | 1.6 | 0.2 | 1.3 | 2.3 |
| **(T:Hg$^{2+}$:T)$_{13}$** (or *T13*) | | | | | | | | | | | | | | | |
| 1.66 | 0.05 | 0.8 | 146.1 | 32.3 | 53.8 | 32.9 | 4.0 | 2.20 | 0.0 | 10.1 | 3.5 | 0.1 | 3.9 | 1.5 | 3.7 |
| **(C:C)$_{13}$** (or Non-metalated *C13*) | | | | | | | | | | | | | | | |
| 0.02 | 1.46 | 51.1 | 37.4 | 48.3 | 24.0 | 9.8 | 17.8 | 33.7 | 1.4 | 0.5 | 1.1 | 0.2 | 0.5 | 1.5 | 1.5 |
| **(C:Ag$^{+}$:C)$_{13}$** (or *C13*) | | | | | | | | | | | | | | | |
| 0.04 | 0.90 | 0.2 | 26.1 | 39.8 | 180.8 | 15.6 | 20.9 | 33.8 | 0.0 | 3.8 | 0.8 | 18.5 | 2.3 | 7.8 | 7.4 |

## C. Spin-flip scattering in the metallo-DNA nanowires:

The presence of spin-orbit coupling (SOC) at the metal centers introduces the possibility of spin-flip scattering in these metallated nanowires. We observed spin-orbit (SO) splitting for an isolated $Hg^{2+}$ ion to be 1.89 eV for HOMO $(5d)$ orbitals and 0.59 eV for LUMO+1 $(6p)$ orbitals. Upon coordination within a nucleobase pair, however, strong orbital hybridization and crystal-field interactions lift the degeneracy of the metal $d$ and $p$ orbitals, even under scalar-relativistic (SR) conditions (*Figure S6, second and third rows*). Incorporating SOC induces non-rigid eigenvalue shifts in these metal-DNA motifs. To quantify SOC-induced spectral variations in the LUMO manifold of a single base pair, we introduce Cumulative Deformation of Levels (CDL): $\left(\sum_i \left|\Delta E_{i,i+1}^{SO} - \Delta E_{i,i+1}^{SR}\right|\right)$, where $\Delta E_{i,i+1} = E_{i+1} - E_i$ represents the energy level spacing between $i^{th}$ and $(i+1)^{th}$ LUMO state within the spin-orbit (SO) and scalar-relativistic (SR) spectra, respectively (see Section 3 of *SI* for additional discussion). At the single base pair level, the CDL for the first two LUMOs, i.e. the SOC-induced change in the LUMO state, is 6.70 meV and 0.25 meV for (T:$Hg^{2+}$:T) and (C:$Ag^{+}$:C) motifs respectively. Extending this analysis across the frontier 13 LUMO states of metallo-DNA nanowire (as these eigenstates have a large metal contributions, see Figure S3) yields a CDL of 19.51 meV for *T13* and 10.50 meV for *C13*. The reduction in energy level splitting relative to the isolated ion stems from orbital hybridization, which reduces the metal contributions to the unoccupied frontier states (Figure S6, fourth and fifth rows). Additionally, the effect of SOC can also be manifested as the difference of SO ($E^{SO}$) and SR ($E^{SR}$) energies $\left(\sum_i \left|E_i^{SO} - E_i^{SR}\right|\right)$, which turns out to be 125.72 meV and 24.29 meV for the first 13 LUMO states of *T13* and *C13*. For a single base pair, the differences are 13.39 meV and 0.79 meV for mercury and silver coordinated motifs respectively. To our knowledge, we have found for the *first time* that SOC remains capable of inducing spin-flip scattering in these nanowires. This effect is reflected in the non-quantized expectation values of the spin operator along the z-axis ($\langle S_Z \rangle$), as shown in Figure S7. To find out if this can induce substantial spin-flip scattering in short nanowires, we considered the nanowire connected to ferromagnetic contacts, which are modeled phenomenologically with coupling constants given by $\Gamma_{L/R} = \Gamma_0(1 \pm p_{L/R})$ for the spin-up and spin-down channels, respectively. Here, $\Gamma_0 = 1\ eV$ and $p_{L/R}$ denotes spin polarization of the left $(L)$ and right $(R)$ contacts which takes only values of (+1) or (-1) in our ferromagnetic contacts. While real ferromagnets exhibit energy-dependent $d$-band features, adopting a wide-band limit helps to ensure that in our calculations, the spin dependence stem from the intrinsic molecular Hamiltonian rather than influenced by the density of states features in the electrode[15,59–63]. The spin-resolved conductances of 13-bp-long metallo-DNA nanowires are presented in Figure 3(b, c). In both systems, the presence of SOC at the metal centers generates finite spin-flip conductances $(T_{\uparrow\downarrow} = T_{\downarrow\uparrow})$. However, the magnitude of the spin-flip conductance differs markedly between the two nanowires. In *T13*, the spin-flip conductance is comparable to the spin-conserving $(T_{\uparrow\uparrow} = T_{\downarrow\downarrow})$ conductance throughout the LUMO manifold (Figure 3(b)), whereas in *C13*, it remains, on average, nearly two orders of magnitude smaller (Figure 3(c)). We have also calculated the spin-resolved conductances for *T13* and *C13* as function of contact coupling strengths, as shown in Figure S8. To further elucidate the role of SOC, Figures 3(d, e) present the average spin-resolved conductance, evaluated over the broadened LUMO band, as a function of nanowire length. The metallo-DNA nanowires show only a modest reduction in conductance before approaching a nearly saturated value at longer length scales (Figure 3(d, e)). This behavior is consistent with the broad metal-assisted conduction bands in Figure 2(a, c), which promote delocalization and suppress the strong localization commonly observed in canonical DNA. The average conductance was obtained over the lowest band in transmission which corresponds to (LUMO to LUMO+2.5 eV) for *C13* and (LUMO to LUMO+1.2 eV) for *T13*. The corresponding Fermi-energy-dependent conductance profiles for all nanowire lengths are provided in Figure S9, while spin-resolved current-voltage characteristics are provided in Figure S10. A comparison between the two nanowires reveals distinct trends in their spin-conserving and spin-flip transport channels. The *C13* nanowire exhibits a higher pure-spin conductance, whereas the *T13* nanowire shows a substantially larger spin-flip conductance. The enhanced pure-spin conductance in *C13* can be traced to its broadened LUMO-derived transport manifold, arising from the close energy spacing of the metal-induced unoccupied states, which promotes efficient electronic delocalization and charge transport along the nanowire (see Figure 2(c) and Table 1). Consistent with this picture, the spin-flip conductance in *C13* remains more than an order of magnitude smaller than the spin-conserving conductance across all length scales, whereas in *T13* the spin-flip and spin-conserving conductance become comparable in magnitude. For *C13*, the relatively weak SOC produces only a small spin rotation per hopping site. Consequently, spin-flip transmission accumulates gradually as the carrier traverses successive $Ag^+$-mediated base pairs, leading to a progressive increase in the spin-flip conductance with nanowire length. In *T13*, stronger *Hg*-induced SOC drives the system into a regime of spin mixing, where substantial spin-flip transmission is generated within only a few hopping sites. As a result, large spin-flip conductance is observed even in relatively short nanowires.

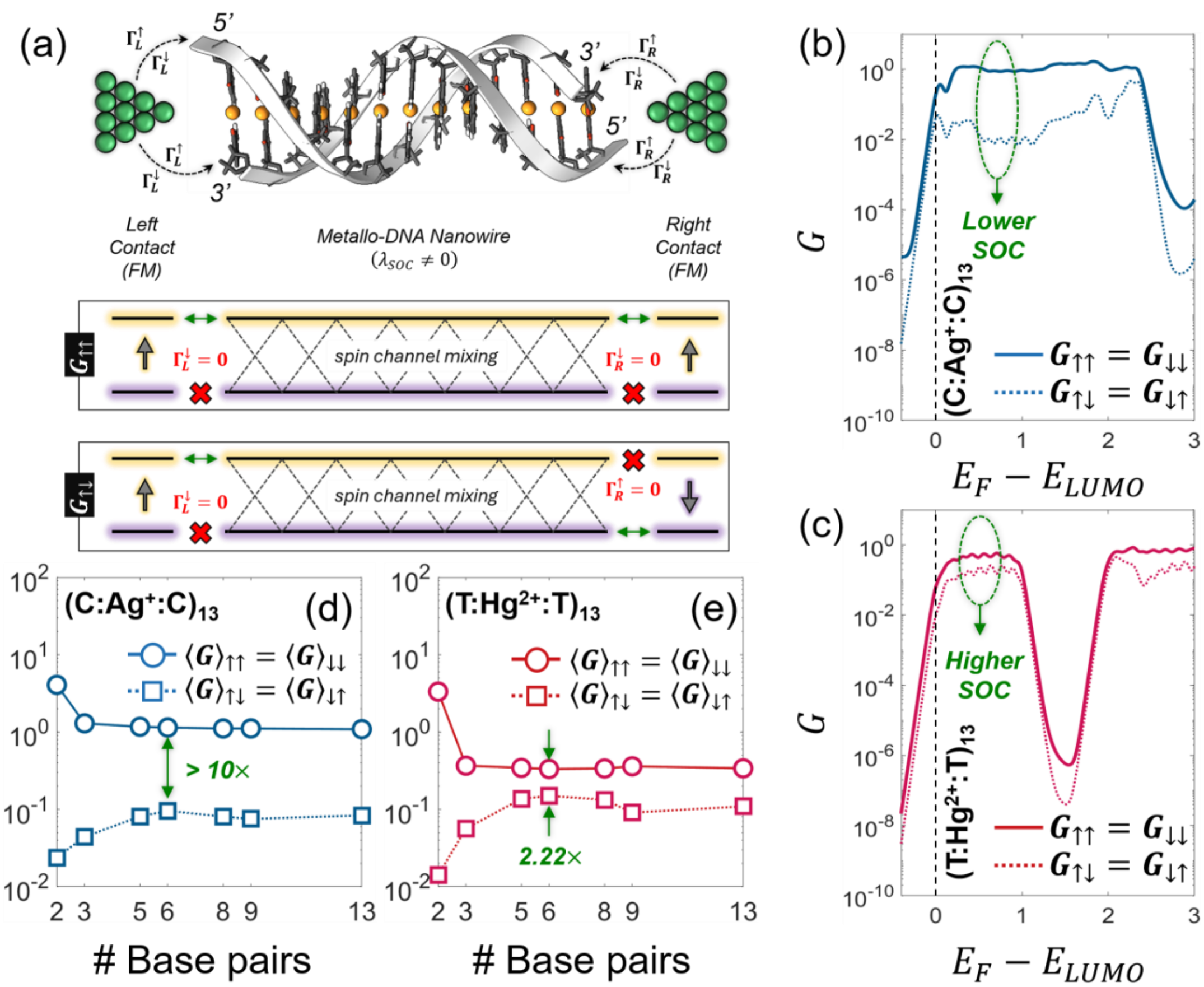

**Figure 3:** (a) (*top panel*) Schematic illustrating the computational setup for spin-resolved conductance. A metallo-DNA nanowire is connected to two ferromagnetic (FM) contacts. The left and right contacts couple to the terminal base pairs through spin-dependent broadening $\Gamma_{L/R}^{\uparrow}$ and $\Gamma_{L/R}^{\downarrow}$ to the *up-spin* and *down-spin* channels, respectively. (*bottom* panel) Pure-spin and spin-flip conductances are obtained by injecting and extracting electrons through the same-spin and opposite-spin channels at the two contacts. Time-reversal symmetry ensures $G_{\uparrow\uparrow} = G_{\downarrow\downarrow}$ and $G_{\uparrow\downarrow} = G_{\downarrow\uparrow}$. The *up-spin* and *down-spin* channels in both contacts and nanowire are represented by *yellow* and *purple* lines. Spin-resolved conductances of 13-bp long (b) $(C:Ag^+:C)_{13}$ and (c) $(T:Hg^{2+}:T)_{13}$ nanowires. Pure-spin ($G_{\uparrow\uparrow} = G_{\downarrow\downarrow}$) and spin-flip ($G_{\uparrow\downarrow} = G_{\downarrow\uparrow}$) conductances are represented by *solid* and *dotted* lines, respectively. The larger spin-flip transmission observed in $(T:Hg^{2+}:T)_{13}$ indicates stronger spin-orbit coupling. Panels (d) and (e) show the *average* pure-spin and spin-flip conductances, evaluated over the corresponding LUMO band, as a function of nanowire length with building blocks of $(C:Ag^+:C)$ and $(T:Hg^{2+}:T)$ metalated base pairs, respectively. The left and right contacts are attached to base pairs '1' and '$N$' for an $N$ base-pair-long nanowire. Here, the conductance ($G$) is in units of $q^2/h$. In the main text, $(T:Hg^{2+}:T)_{13}$ and $(C:Ag^+:C)_{13}$ have also been referred to as *T13* and *C13*.

## Conclusion

In this work, we investigate how heavy-metal ion coordination modifies the electronic structure and spin-dependent transport properties of metallo-DNA nanowires using a relativistic two-component framework. Our central finding is that heavy-metal coordination can generate strong spin-flip scattering in DNA while simultaneously altering its fundamental electronic properties. We find that metal coordination transforms a weakly coupled DNA system into a tunable quasi-one-dimensional conductor. The metal centers substantially enhance interlayer coupling along the helical axis, as evidenced by greater delocalization of frontier orbitals and broader transmission bands (Figure 2). These signatures indicate reduced backscattering and improved tolerance to structural fluctuations. We compare $Ag^{+}$- and $Hg^{2+}$-coordinated systems, whose metal centers differ in the strength of their relativistic effects. $Hg^{2+}$ coordination generates states significantly deeper within the bandgap than $Ag^{+}$ coordination, leading to a greater reduction in the effective bandgap (Figure 1(a, bottom)). This observation may be relevant to the larger increase in conductance observed experimentally upon incorporating $T{:}Hg^{2+}{:}T$ pairs compared with $C{:}Ag^{+}$:C pairs in break-junction experiments.[64,65] In the $Hg^{2+}$-coordinated nanowire, we further identify a sequence of quantized states with a clear oscillating nodal pattern spanning the lowest unoccupied energy levels (Figure 1(d)), in contrast to the $Ag^{+}$-coordinated nanowire (Figure S4). We attribute this distinction to stronger $Hg$-$Hg$ interaction relative to $Ag$-$Ag$ , as reflected in the highly delocalized low-lying unoccupied states. We further predict that relativistic corrections significantly shift these states and increase their metal character. While spin–orbit splitting is suppressed near the transport-relevant frontier states by crystal-field effects and ligand hybridization, it still manifests as a non-rigid deformation of the energy spectrum, quantified by cumulative deformation of levels (CDL). Notably, these CDL values are comparable to the next-nearest-neighbor hopping energies (Table 1), suggesting that SOC is sufficiently strong to drive substantial spin mixing, as evidenced by non-quantized spin expectation values (see $\langle S_z \rangle$ in Figure S7).

A key result is the emergence of pronounced spin-flip transport upon heavy-metal incorporation. Although spin-flip processes have been discussed in theoretical models, explicit quantitative estimates of spin-flip conductance in native DNA remain scarce, primarily because of the weak intrinsic SOC of its light constituent atoms[66–69]. In our work, metallo-DNA nanowires offer an opportunity to achieve and quantify substantial spin-flip conductance arising from strong relativistic effects at the heavy-metal centers. We predict that, even at length scales typical of transport experiments, spin-flip conductance in the $Hg^{2+}$-coordinated nanowire can approach the spin-conserving conductances across much of the low-lying LUMO energy band. This demonstrates that heavy-metal centers can act as efficient molecular spin scatterers (Figure 3(c)). In contrast, spin-flip conductance is substantially smaller in the $Ag^{+}$-coordinated nanowire (Figure 3(b)). In both systems, spin-flip transport originates from the spin character of the delocalized conducting states. Heavy-metal orbitals mix opposite-spin components even in the frontier states, while strong electronic coupling along the helical axis allows these states to propagate through the nanowire. Consequently, spin conversion occurs over a broad energy window rather than being restricted to a few resonant states. The metal-induced bands also support spin-dependent transport over extended molecular lengths, with only a modest reduction in conductance as the nanowire length increases (Figure 3(d, e)). This weak length dependence reflects the enhanced coupling and spatial delocalization of the metal-derived frontier states. With dephasing included, however, we expect the conductance to decrease as $1/L$ in the limit of large $L$, where $L$ denotes the DNA length.

Overall, our findings establish metallo-DNA nanowires as tunable quasi-one-dimensional conductors whose transport properties are governed by metal type, hybridization strength, and relativistic effects. These results demonstrate that heavy-metal coordination can be used not only to enhance charge transport but also to engineer spin scattering within a self-assembled molecular wire. The combination of DNA self-assembly, heavy-metal orbitals, and tunable interlayer coupling may enable the realization of nanoscale quantum wires, spin filters, and programmable molecular interconnects.

## Supporting Information

Atomic representations of modeled structures (Figure S1); Comparison of Non-relativistic (NR) and Scalar-relativistic (SR) Calculations. (Figure S2); Metal contributions in the metallo-DNA nanowires (Figure S3); Energy levels, DOS and Probability Density of $Ag^{+}$-coordinated metallo-DNA nanowires (Figure S4); Linear-response conductance in logarithmic scale (Figure S5); Contribution of $s, p,$ and $d$ orbitals of metal in different metal-DNA configurations (Figure S6); Spin expectation (Figure S7); Spin-resolved conductance as a function of contact coupling strength (Figure S8); Spin-resolved conductance at different length scales (Figure S9); Spin-Resolved Current-Voltage (I-V) Characteristics (Figure S10); Benchmarking Spin-orbit splitting with GKS calculations (Table S1); Additional discussion on Cumulative Deformation of Levels (CDL).

## Acknowledgements

We acknowledge support for NSF GCR-2317843, GR-046708 and NSF 2328815. We thank Prof. Arindam K Das for discussion, Prof. Antia S Botana for feedback on the manuscript, and the HYAK supercomputing facility at the University of Washington.

## Author Contributions

MPA conceptualized the research problem. The framework to evaluate the extent of spin-orbit coupling was formalized by AD and MPA. The results were analyzed and interpreted by AD and MPA. All calculations were performed AD, and the computational code was developed by AD. Data curation and visualization were done by AD. First draft was written by AD and MPA. All authors reviewed and edited the manuscript. MPA supervised the research work.

## References

(1) Murphy, C. J.; Arkin, M. R.; Jenkins, Y.; Ghatlia, N. D.; Bossmann, S. H.; Turro, N. J.; Barton, J. K. Long-Range Photoinduced Electron Transfer Through a DNA Helix. *Science* **1993**, *262* (5136), 1025–1029. https://doi.org/10.1126/science.7802858

(2) Arkin, M. R.; Stemp, E. D. A.; Holmlin, R. E.; Barton, J. K.; Hörmann, A.; Olson, E. J. C.; Barbara, P. F. Rates of DNA-Mediated Electron Transfer Between Metallointercalators. *Science* **1996**, *273* (5274), 475–480. https://doi.org/10.1126/science.273.5274.475

(3) Fink, H.-W.; Schönenberger, C. Electrical Conduction through DNA Molecules. *Nature* **1999**, *398* (6726), 407–410. https://doi.org/10.1038/18855

(4) Alexandre, S. S.; Artacho, E.; Soler, J. M.; Chacham, H. Small Polarons in Dry DNA. *Phys. Rev. Lett.* **2003**, *91* (10), 108105. https://doi.org/10.1103/PhysRevLett.91.108105

(5) Yoo, K.-H.; Ha, D. H.; Lee, J.-O.; Park, J. W.; Kim, J.; Kim, J. J.; Lee, H.-Y.; Kawai, T.; Choi, H. Y. Electrical Conduction through Poly(dA)-Poly(dT) and Poly(dG)-Poly(dC) DNA Molecules. *Phys. Rev. Lett.* **2001**, *87* (19), 198102. https://doi.org/10.1103/PhysRevLett.87.198102

(6) Porath, D.; Bezryadin, A.; de Vries, S.; Dekker, C. Direct Measurement of Electrical Transport through DNA Molecules. *Nature* **2000**, *403* (6770), 635–638. https://doi.org/10.1038/35001029

(7) Kasumov, A. Yu.; Kociak, M.; Guéron, S.; Reulet, B.; Volkov, V. T.; Klinov, D. V.; Bouchiat, H. Proximity-Induced Superconductivity in DNA. *Science* **2001**, *291* (5502), 280–282. https://doi.org/10.1126/science.291.5502.280

(8) Adessi, Ch.; Walch, S.; Anantram, M. P. Environment and Structure Influence on DNA Conduction. *Phys. Rev. B* **2003**, *67* (8), 081405. https://doi.org/10.1103/PhysRevB.67.081405

(9) Tan, B.; Hodak, M.; Lu, W.; Bernholc, J. Charge Transport in DNA Nanowires Connected to Carbon Nanotubes. *Phys. Rev. B* **2015**, *92* (7), 075429. https://doi.org/10.1103/PhysRevB.92.075429

(10) Roche, S. Long Range Correlations in DNA: Scaling Properties and Charge Transfer Efficiency. *Phys. Rev. Lett.* **2003**, *91* (22). https://doi.org/10.1103/PhysRevLett.91.228101

(11) Jortner, J.; Bixon, M.; Langenbacher, T.; Michel-Beyerle, M. E. Charge Transfer and Transport in DNA. *Proceedings of the National Academy of Sciences* **1998**, *95* (22), 12759–12765. https://doi.org/10.1073/pnas.95.22.12759

(12) Livshits, G. I.; Stern, A.; Rotem, D.; Borovok, N.; Eidelshtein, G.; Migliore, A.; Penzo, E.; Wind, S. J.; Di Felice, R.; Skourtis, S. S.; Cuevas, J. C.; Gurevich, L.; Kotlyar, A. B.; Porath, D. Long-Range Charge Transport in Single G-Quadruplex DNA Molecules. *Nature Nanotech* **2014**, *9* (12), 1040–1046. https://doi.org/10.1038/nnano.2014.246

(13) Xu, M.; Chen, Y. Enhancement of Chirality Induced Spin Selectivity by Strong Electron Correlations. *Phys. Rev. B* **2024**, *110* (23), 235145. https://doi.org/10.1103/PhysRevB.110.235145

(14) Du, G.-F.; Fu, H.-H.; Wu, R. Vibration-Enhanced Spin-Selective Transport of Electrons in the DNA Double Helix. *Phys. Rev. B* **2020**, *102* (3), 035431. https://doi.org/10.1103/PhysRevB.102.035431

(15) Hu, P.-J.; Wang, S.-X.; Gao, X.-H.; Zhang, Y.-Y.; Fang, T.-F.; Guo, A.-M.; Sun, Q.-F. Spin-Dependent Electron Transport along Hairpinlike DNA Molecules. *Phys. Rev. B* **2020**, *102* (19), 195406. https://doi.org/10.1103/PhysRevB.102.195406

(16) Patil, S. R.; Chawda, V.; Qi, J.; Anantram, M. P.; Sinha, N. Charge Transport through DNA Based Electronic Barriers. *AIP Conf. Proc.* **2018**, *1953* (1), 140148. https://doi.org/10.1063/1.5033323

(17) Mallajosyula, S. S.; Lin, J. C.; Cox, D. L.; Pati, S. K.; Singh, R. R. P. Sequence Dependent Electron Transport in Wet DNA: Ab Initio and Molecular Dynamics Studies. *Phys. Rev. Lett.* **2008**, *101* (17), 176805. https://doi.org/10.1103/PhysRevLett.101.176805

(18) Song, Y.; Gao, Y.; Fang, H. Unexpected Large Charge Transfer Rate Mediated by Adenine in Twisted DNA Structure. *Phys. Rev. E* **2024**, *109* (6), 064412. https://doi.org/10.1103/PhysRevE.109.064412

(19) Patil, S. R.; Mohammad, H.; Chawda, V.; Sinha, N.; Singh, R. K.; Qi, J.; Anantram, M. P. Quantum Transport in DNA Heterostructures: Implications for Nanoelectronics. *ACS Appl. Nano Mater.* **2021**, *4* (10), 10029–10037. https://doi.org/10.1021/acsanm.1c01087
(20) Qi, J.; Edirisinghe, N.; Rabbani, M. G.; Anantram, M. P. Unified Model for Conductance through DNA with the Landauer-Büttiker Formalism. *Physical Review B - Condensed Matter and Materials Physics* **2013**, *87* (8), 1–10. https://doi.org/10.1103/PhysRevB.87.085404
(21) Voityuk, A. A.; Rösch, N.; Bixon, M.; Jortner, J. Electronic Coupling for Charge Transfer and Transport in DNA. *J. Phys. Chem. B* **2000**, *104* (41), 9740–9745. https://doi.org/10.1021/jp001109w
(22) Voityuk, A. A.; Jortner, J.; Bixon, M.; Rösch, N. Electronic Coupling between Watson–Crick Pairs for Hole Transfer and Transport in Desoxyribonucleic Acid. *J. Chem. Phys.* **2001**, *114* (13), 5614–5620. https://doi.org/10.1063/1.1352035
(23) Mehrez, H.; Anantram, M. P. Interbase Electronic Coupling for Transport through DNA. *Phys. Rev. B* **2005**, *71* (11), 115405. https://doi.org/10.1103/PhysRevB.71.115405
(24) Kundu, S.; Karmakar, S. N. Localization Phenomena in a DNA Double-Helix Structure: A Twisted Ladder Model. *Phys. Rev. E* **2014**, *89* (3), 032719. https://doi.org/10.1103/PhysRevE.89.032719
(25) Vecchioni, S.; Lu, B.; Livernois, W.; Ohayon, Y. P.; Yoder, J. B.; Yang, C.-F.; Woloszyn, K.; Bernfeld, W.; Anantram, M. P.; Canary, J. W.; Hendrickson, W. A.; Rothschild, L. J.; Mao, C.; Wind, S. J.; Seeman, N. C.; Sha, R. Metal-Mediated DNA Nanotechnology in 3D: Structural Library by Templated Diffraction. *Advanced Materials* **2023**, *35* (29), 2210938. https://doi.org/10.1002/adma.202210938
(26) Vecchioni, S.; Capece, M. C.; Toomey, E.; Nguyen, L.; Ray, A.; Greenberg, A.; Fujishima, K.; Urbina, J.; Paulino-Lima, I. G.; Pinheiro, V.; Shih, J.; Wessel, G.; Wind, S. J.; Rothschild, L. Construction and Characterization of Metal Ion-Containing DNA Nanowires for Synthetic Biology and Nanotechnology. *Sci Rep* **2019**, *9* (1), 6942. https://doi.org/10.1038/s41598-019-43316-1
(27) Bethur, E.; Guha, R.; Zhao, Z.; Katz, B. B.; Ashby, P. D.; Zeng, H.; Copp, S. M. Formation and Nanomechanical Properties of Silver-Mediated Guanine DNA Duplexes in Aqueous Solution. *ACS Nano* **2024**, *18* (4), 3002–3010. https://doi.org/10.1021/acsnano.3c08008
(28) Kondo, J.; Tada, Y.; Dairaku, T.; Hattori, Y.; Saneyoshi, H.; Ono, A.; Tanaka, Y. A Metallo-DNA Nanowire with Uninterrupted One-Dimensional Silver Array. *Nature Chemistry 2017 9:10* **2017**, *9* (10), 956–960. https://doi.org/10.1038/nchem.2808
(29) Ono, A.; Kanazawa, H.; Ito, H.; Goto, M.; Nakamura, K.; Saneyoshi, H.; Kondo, J. A Novel DNA Helical Wire Containing HgII-Mediated T:T and T:G Pairs. *Angewandte Chemie International Edition* **2019**, *58* (47), 16835–16838. https://doi.org/10.1002/anie.201910029
(30) Kondo, J.; Yamada, T.; Hirose, C.; Okamoto, I.; Tanaka, Y.; Ono, A. Crystal Structure of Metallo DNA Duplex Containing Consecutive Watson–Crick-like T–HgII–T Base Pairs. *Angewandte Chemie International Edition* **2014**, *53* (9), 2385–2388. https://doi.org/10.1002/anie.201309066
(31) Corfdir, P.; Li, H.; Marquardt, O.; Gao, G.; Molas, M. R.; Zettler, J. K.; van Treeck, D.; Flissikowski, T.; Potemski, M.; Draxl, C.; Trampert, A.; Fernández-Garrido, S.; Grahn, H. T.; Brandt, O. Crystal-Phase Quantum Wires: One-Dimensional Heterostructures with Atomically Flat Interfaces. *Nano Lett.* **2018**, *18* (1), 247–254. https://doi.org/10.1021/acs.nanolett.7b03997
(32) Asaba, T.; Peng, L.; Ono, T.; Akutagawa, S.; Tanaka, I.; Murayama, H.; Suetsugu, S.; Razpopov, A.; Kasahara, Y.; Terashima, T.; Kohsaka, Y.; Shibauchi, T.; Ichikawa, M.; Valentí, R.; Sasa, S.; Matsuda, Y. Growth of Self-Integrated Atomic Quantum Wires and Junctions of a Mott Semiconductor. *Science Advances* **2023**, *9* (18), eabq5561. https://doi.org/10.1126/sciadv.abq5561
(33) De, A.; Lu, B.; Ohayon, Y. P.; Woloszyn, K.; Livernois, W.; Perren, L.; Yang, C.; Mao, C.; Botana, A. S.; Hihath, J.; Canary, J. W.; Sha, R.; Anantram, M. p.; Vecchioni, S. Transmetalation for DNA-Based Molecular Electronics. *Small* **2025**, *21* (25), 2411518. https://doi.org/10.1002/smll.202411518
(34) Figgen, D.; Rauhut, G.; Dolg, M.; Stoll, H. Energy-Consistent Pseudopotentials for Group 11 and 12 Atoms: Adjustment to Multi-Configuration Dirac–Hartree–Fock Data. *Chemical Physics* **2005**, *311* (1), 227–244. https://doi.org/10.1016/j.chemphys.2004.10.005
(35) Hemstreet, L. A.; Fong, C. Y.; Nelson, J. S. First-Principles Calculations of Spin-Orbit Splittings in Solids Using Nonlocal Separable Pseudopotentials. *Phys. Rev. B* **1993**, *47* (8), 4238–4243. https://doi.org/10.1103/PhysRevB.47.4238
(36) Cuadrado, R.; Cerdá, J. I. Fully Relativistic Pseudopotential Formalism under an Atomic Orbital Basis: Spin–Orbit Splittings and Magnetic Anisotropies. *J. Phys.: Condens. Matter* **2012**, *24* (8), 086005. https://doi.org/10.1088/0953-8984/24/8/086005
(37) Kleinman, L. Relativistic Norm-Conserving Pseudopotential. *Phys. Rev. B* **1980**, *21* (6), 2630–2631. https://doi.org/10.1103/PhysRevB.21.2630
(38) Bachelet, G. B. Relativistic Norm-Conserving Pseudopotentials. *Phys. Rev. B* **1982**, *25* (4), 2103–2108. https://doi.org/10.1103/PhysRevB.25.2103
(39) Ullrich, C. A. Density-Functional Theory for Systems with Noncollinear Spin: Orbital-Dependent Exchange-Correlation Functionals and Their Application to the Hubbard Dimer. *Phys. Rev. B* **2018**, *98* (3), 035140. https://doi.org/10.1103/PhysRevB.98.035140
(40) Case, D. A.; Aktulga, H. M.; Belfon, K.; Cerutti, D. S.; Cisneros, G. A.; Cruzeiro, V. W. D.; Forouzesh, N.; Giese, T. J.; Götz, A. W.; Gohlke, H.; Izadi, S.; Kasavajhala, K.; Kaymak, M. C.; King, E.; Kurtzman, T.; Lee, T.-S.; Li, P.; Liu, J.; Luchko, T.; Luo, R.; Manathunga, M.; Machado, M. R.; Nguyen, H. M.; O'Hearn, K. A.; Onufriev, A. V.; Pan, F.; Pantano, S.; Qi, R.; Rahnamoun, A.; Risheh, A.; Schott-Verdugo, S.; Shajan, A.; Swails, J.; Wang, J.; Wei, H.; Wu, X.; Wu, Y.; Zhang, S.; Zhao, S.;

Zhu, Q.; Cheatham, T. E. I.; Roe, D. R.; Roitberg, A.; Simmerling, C.; York, D. M.; Nagan, M. C.; Merz, K. M. Jr. AmberTools. *J. Chem. Inf. Model.* **2023**, *63* (20), 6183–6191. https://doi.org/10.1021/acs.jcim.3c01153
(41) Atsugi, T.; Ono, A.; Tasaka, M.; Eguchi, N.; Fujiwara, S.; Kondo, J. A Novel AgI-DNA Rod Comprising a One-Dimensional Array of 11 Silver Ions within a Double Helical Structure. *Angewandte Chemie International Edition* **2022**, *61* (31), e202204798. https://doi.org/10.1002/anie.202204798
(42) Kondo, J.; Yamada, T.; Hirose, C.; Okamoto, I.; Tanaka, Y.; Ono, A. Crystal Structure of Metallo DNA Duplex Containing Consecutive Watson–Crick-like T–HgII–T Base Pairs. *Angewandte Chemie International Edition* **2014**, *53* (9), 2385–2388. https://doi.org/10.1002/anie.201309066
(43) Livernois, W.; Alolaiyan, O.; De, A.; Anantram, M. P. Scalable Force Fields for Metal-Mediated DNA Nanostructures. *J. Chem. Theory Comput.* **2026**, *22* (4), 2000–2012. https://doi.org/10.1021/acs.jctc.5c01426. For the Hg-N and Ag-N distances, the reader is refered to the following files of the JCTC paper: "SingleCHgT_solv_mdmin.pdb", "SingleCHgT_FF_mineq.pdb", "SingleCAg2T_solv_mdmin.pdb", and "SingleCAg2T_FF_mineq.pdb"
(44) Gaussian Development Version, 2023.
(45) Wang, Y.; Demir, B.; Mohammad, H.; Oren, E. E.; Anantram, M. P. Computational Study of the Role of Counterions and Solvent Dielectric in Determining the Conductance of B-DNA. *Phys. Rev. E* **2023**, *107* (4), 044404. https://doi.org/10.1103/PhysRevE.107.044404
(46) Li, Y.; Artés, J. M.; Demir, B.; Gokce, S.; Mohammad, H. M.; Alangari, M.; Anantram, M. P.; Oren, E. E.; Hihath, J. Detection and Identification of Genetic Material via Single-Molecule Conductance. *Nature Nanotechnology 2018 13:12* **2018**, *13* (12), 1167–1173. https://doi.org/10.1038/s41565-018-0285-x
(47) Artés, J. M.; Li, Y.; Qi, J.; Anantram, M. P.; Hihath, J. Conformational Gating of DNA Conductance. *Nat Commun* **2015**, *6* (1), 8870. https://doi.org/10.1038/ncomms9870
(48) Creutzberg, J.; Hedegård, E. D. Investigating the Influence of Relativistic Effects on Absorption Spectra for Platinum Complexes with Light-Activated Activity against Cancer Cells. *Phys. Chem. Chem. Phys.* **2020**, *22* (46), 27013–27023. https://doi.org/10.1039/D0CP05143H
(49) Bjornsson, R.; Bühl, M. Electric Field Gradients of Transition Metal Complexes from Density Functional Theory: Assessment of Functionals, Geometries and Basis Sets. *Dalton Trans.* **2010**, *39* (22), 5319–5324. https://doi.org/10.1039/C001017K
(50) Ito, S.; Koyasu, K.; Tsukuda, T. Structural Origin of Spin-Orbit Splitting of 1P Orbitals of Icosahedral Au13 Superatom Protected by Ligands. *Angewandte Chemie International Edition* **2025**, *64* (30), e202508151. https://doi.org/10.1002/anie.202508151
(51) Arpan De; Hashem Mohammad; Yiren Wang; Rajkumar Kubendran; Arindam K. Das; M. P. Anantram. Performance Analysis of DNA Crossbar Arrays for High-Density Memory Storage Applications. *Scientific Reports*. https://doi.org/10.1038/s41598-023-33004-6
(52) Mohammad, H.; Anantram, M. P. Charge Transport through DNA with Energy-Dependent Decoherence. *Phys. Rev. E* **2023**, *108* (4), 044403. https://doi.org/10.1103/PhysRevE.108.044403
(53) Bhai, S.; Ganguly, B. Role of the Backbone of Nucleic Acids in the Stability of Hg2+-Mediated Canonical Base Pairs and Thymine–Thymine Mispair: A DFT Study. *RSC Adv.* **2020**, *10* (67), 40969–40982. https://doi.org/10.1039/d0ra07526d
(54) Naskar, S.; Guha, R.; Müller, J. Metal-Modified Nucleic Acids: Metal-Mediated Base Pairs, Triples, and Tetrads. *Angewandte Chemie International Edition* **2020**, *59* (4), 1397–1406. https://doi.org/10.1002/anie.201905913
(55) Clever, G. H.; Shionoya, M. Alternative DNA Base Pairing through Metal Coordination. In *Interplay between Metal Ions and Nucleic Acids*; Sigel, A., Sigel, H., Sigel, R. K. O., Eds.; Springer Netherlands: Dordrecht, 2012; pp 269–294. https://doi.org/10.1007/978-94-007-2172-2_10
(56) Gonzàlez-Rosell, A.; Copp, S. M. An Atom-Precise Understanding of DNA-Stabilized Silver Nanoclusters. *Acc. Chem. Res.* **2024**, *57* (15), 2117–2129. https://doi.org/10.1021/acs.accounts.4c00256
(57) Swasey, S. M.; Rosu, F.; Copp, S. M.; Gabelica, V.; Gwinn, E. G. Parallel Guanine Duplex and Cytosine Duplex DNA with Uninterrupted Spines of AgI-Mediated Base Pairs. *J. Phys. Chem. Lett.* **2018**, *9* (22), 6605–6610. https://doi.org/10.1021/acs.jpclett.8b02851
(58) Javornik, U.; Pérez-Romero, A.; López-Chamorro, C.; Smith, R. M.; Dobado, J. A.; Palacios, O.; Bera, M. K.; Nyman, M.; Plavec, J.; Galindo, M. A. Unveiling the Solution Structure of a DNA Duplex with Continuous Silver-Modified Watson-Crick Base Pairs. *Nat Commun* **2024**, *15* (1), 7763. https://doi.org/10.1038/s41467-024-51876-8
(59) Weymann, I.; Wójcik, K. P. Andreev Transport in a Correlated Ferromagnet-Quantum-Dot-Superconductor Device. *Phys. Rev. B* **2015**, *92* (24), 245307. https://doi.org/10.1103/PhysRevB.92.245307
(60) Martinek, J.; Utsumi, Y.; Imamura, H.; Barnaś, J.; Maekawa, S.; König, J.; Schön, G. Kondo Effect in Quantum Dots Coupled to Ferromagnetic Leads. *Phys. Rev. Lett.* **2003**, *91* (12), 127203. https://doi.org/10.1103/PhysRevLett.91.127203
(61) Sindel, M.; Borda, L.; Martinek, J.; Bulla, R.; König, J.; Schön, G.; Maekawa, S.; von Delft, J. Kondo Quantum Dot Coupled to Ferromagnetic Leads: Numerical Renormalization Group Study. *Phys. Rev. B* **2007**, *76* (4), 045321. https://doi.org/10.1103/PhysRevB.76.045321
(62) Weymann, I. Finite-Temperature Spintronic Transport through Kondo Quantum Dots: Numerical Renormalization Group Study. *Phys. Rev. B* **2011**, *83* (11), 113306. https://doi.org/10.1103/PhysRevB.83.113306
(63) Fransson, J. Vibrational Origin of Exchange Splitting and ’ ’chiral-Induced Spin Selectivity. *Phys. Rev. B* **2020**, *102* (23), 235416. https://doi.org/10.1103/PhysRevB.102.235416

(64) Toomey, E.; Xu, J.; Vecchioni, S.; Rothschild, L.; Wind, S.; Fernandes, G. E. Comparison of Canonical versus Silver(I)-Mediated Base-Pairing on Single Molecule Conductance in Polycytosine dsDNA. *J. Phys. Chem. C* **2016**, *120* (14), 7804–7809. https://doi.org/10.1021/acs.jpcc.5b11968

(65) Yamada, R.; Nomura ,Issei; Yamaguchi ,Yuki; Matsuda ,Yosuke; Hattori ,Yoshikazu; Tada ,Hirokazu; Ono ,Akira; and Tanaka, Y. Electrical Conductance Measurement of HgII-Mediated DNA Duplex in Buffered Aqueous Solution. *Nucleosides, Nucleotides & Nucleic Acids* **2020**, *39* (8), 1083–1087. https://doi.org/10.1080/15257770.2020.1755044

(66) Kang, D. W.; Hao, X. P.; Li, X. Z.; Li, L. B.; Xie, S. J. Spin Polarized Current through Cu-DNA Modulated by a Gate Voltage. *Appl. Phys. Lett.* **2013**, *102* (7), 072410. https://doi.org/10.1063/1.4793479

(67) Utsumi, Y.; Entin-Wohlman, O.; Aharony, A. Spin Selectivity through Time-Reversal Symmetric Helical Junctions. *Phys. Rev. B* **2020**, *102* (3), 035445. https://doi.org/10.1103/PhysRevB.102.035445

(68) Stemer, D. M.; Abendroth, J. M.; Cheung, K. M.; Ye, M.; El Hadri, M. S.; Fullerton, E. E.; Weiss, P. S. Differential Charging in Photoemission from Mercurated DNA Monolayers on Ferromagnetic Films. *Nano Lett.* **2020**, *20* (2), 1218–1225. https://doi.org/10.1021/acs.nanolett.9b04622

(69) Roy, A. K.; Speyer, J. L.; Bartell, L.; Neuhauser, D. Spin-Birefringence in Molecular Currents: Tellurium and Gold Complexes. *Chemical Physics Letters* **2010**, *484* (4), 104–109. https://doi.org/10.1016/j.cplett.2009.12.001